\documentclass[preprint]{aastex62}
\usepackage{array}
\usepackage{color}

\newcommand{\CC}{{\tt C1.0+/0/24}}
\newcommand{\MM}{{\tt M1.0+/0/24}}
\newcommand{\Cp}{{\tt C1.0+}}
\newcommand{\Mp}{{\tt M1.0+}}
\newcommand{\Xp}{{\tt X1.0+}}

\newcommand{\Co}{{\tt C1.0--C9.9}}
\newcommand{\Mo}{{\tt M1.0--M9.9}}

\received{}
\revised{}
\accepted{}

\submitjournal{ApJSS}

\begin{document}
\title{A Comparison of Flare Forecasting Methods. II. Benchmarks, Metrics and Performance Results for Operational Solar Flare Forecasting Systems}

\correspondingauthor{K.~D.~Leka}
\email{kdleka@isee.nagoya-u.ac.jp, leka@nwra.com}

\author[0000-0003-0026-931X]{K.D. Leka}
\affiliation{Institute for Space-Earth Environmental Research
Nagoya University 
Furo-cho Chikusa-ku 
Nagoya, Aichi 464-8601 JAPAN}
\affiliation{NorthWest Research Associates
3380 Mitchell Lane
Boulder, CO 80301 USA}

\author[0000-0001-9149-6547]{Sung-Hong Park}
\affiliation{Institute for Space-Earth Environmental Research
Nagoya University 
Furo-cho Chikusa-ku 
Nagoya, Aichi 464-8601 JAPAN}

\author[0000-0002-6814-6810]{Kanya Kusano}
\affiliation{Institute for Space-Earth Environmental Research
Nagoya University 
Furo-cho Chikusa-ku 
Nagoya, Aichi 464-8601 JAPAN}

\author{Jesse Andries}
\affiliation{STCE - Royal Observatory of Belgium 
Avenue Circulaire, 3 
B-1180 Brussels BELGIUM}

\author[0000-0003-3571-8728]{Graham Barnes}
\affiliation{NorthWest Research Associates
3380 Mitchell Lane
Boulder, CO 80301 USA}

\author{Suzy Bingham}
\affiliation{Met Office
FitzRoy Road
Exeter, Devon, EX1 3PB, UNITED KINGDOM}

\author[0000-0002-4183-9895]{D. Shaun Bloomfield}
\affiliation{Northumbria University
Newcastle upon Tyne 
NE1 8ST, UNITED KINGDOM}

\author[0000-0002-4830-9352]{Aoife E. McCloskey}
\affiliation{School of Physics,
Trinity College Dublin,
College Green,
Dublin 2, IRELAND}

\author[0000-0001-5307-8045]{Veronique Delouille}
\affiliation{STCE - Royal Observatory of Belgium 
Avenue Circulaire, 3 
B-1180 Brussels BELGIUM}

\author{David Falconer}
\affiliation{NASA/NSSTC 
Mail Code ST13 
320 Sparkman Drive Huntsville, AL 35805, USA}

\author[0000-0001-9745-0400]{Peter T. Gallagher}
\affiliation{School of Cosmic Physics,
Dublin Institute for Advanced Studies,
31 Fitzwilliam Place,
Dublin, D02 XF86, IRELAND}

\author[0000-0001-6913-1330]{Manolis K. Georgoulis}
\affiliation{Department of Physics \& Astronomy 
Georgia State University 
1 Park Place, Rm \#715,
Atlanta, GA 30303, USA}
\affiliation{Academy of Athens
4 Soranou Efesiou Street, 11527 Athens, GREECE}

\author{Yuki Kubo}
\affiliation{National Institute of Information and Communications Technology\\
Space Environment Laboratory\\
4-2-1 Nukuikita Koganei Tokyo 184-8795 JAPAN}

\author{Kangjin Lee}
\affiliation{Meteorological Satellite Ground Segment Development Center\\
Electronics and Telecommunications Research Institute, Daejeon \\
218 Gajeong-ro, Yuseong-gu, Daejeon, 34129, REPUBLIC OF KOREA}
\affiliation{Kyung Hee University \\
1732, Deogyeong-daero, Giheung-gu, Yongin, 17104, REPUBLIC OF KOREA}

\author{Sangwoo Lee}
\affiliation{SELab, Inc. 
150-8, Nonhyeon-ro, Gangnam-gu, Seoul, 06049, REPUBLIC OF KOREA}

\author[0000-0001-5655-9928]{Vasily Lobzin}
\affiliation{Bureau of Meteorology 
Space Weather Services 
PO Box 1386 Haymarket NSW 1240 AUSTRALIA}

\author{JunChul Mun}
\affiliation{Korean Space Weather Center 
198-6, Gwideok-ro, Hallim-eup, Jeju-si, 63025, REPUBLIC OF KOREA}

\author[0000-0002-9378-5315]{Sophie A. Murray}
\affiliation{School of Physics,
Trinity College Dublin,
College Green,
Dublin 2, IRELAND}
\affiliation{School of Cosmic Physics,
Dublin Institute for Advanced Studies,
31 Fitzwilliam Place,
Dublin, D02 XF86, IRELAND}

\author{Tarek A.M. Hamad Nageem}
\affiliation{University of Bradford 
Bradford West Yorkshire BD7 1DP UK}

\author[0000-0002-8637-1130]{Rami Qahwaji}
\affiliation{University of Bradford 
Bradford West Yorkshire BD7 1DP UNITED KINGDOM}

\author{Michael Sharpe}
\affiliation{Met Office
FitzRoy Road
Exeter, Devon, EX1 3PB, UNITED KINGDOM}

\author[0000-0001-8123-4244]{Rob Steenburgh}
\affiliation{NOAA/National Weather Service 
National Centers for Environmental Prediction 
Space Weather Prediction Center, W/NP9 
325 Broadway, Boulder CO 80305 USA}

\author[0000-0002-9176-2697]{Graham Steward}
\affiliation{Bureau of Meteorology 
Space Weather Services 
PO Box 1386 Haymarket NSW 1240 AUSTRALIA}

\author[0000-0002-6290-158X]{Michael Terkildsen}
\affiliation{Bureau of Meteorology 
Space Weather Services 
PO Box 1386 Haymarket NSW 1240 AUSTRALIA}

\begin{abstract}
Solar flares are extremely energetic phenomena in our Solar System. Their
impulsive, often drastic radiative increases, in particular at short
wavelengths, bring immediate impacts that motivate solar physics and space
weather research to understand solar flares to the point of being able
to forecast them. As data and algorithms improve dramatically, questions
must be asked concerning how well the forecasting performs; crucially,
we must ask {\it how} to rigorously measure performance in order to critically
gauge any improvements. Building upon earlier-developed methodology 
\citep[][Paper I]{allclear}, international representatives of regional warning
centers and research facilities assembled in 2017 at the Institute for
Space-Earth Environmental Research, Nagoya University, Japan to -- for
the first time -- directly compare the performance of operational solar
flare forecasting methods. Multiple quantitative evaluation metrics are
employed, with focus and discussion on evaluation methodologies given
the restrictions of operational forecasting. Numerous methods performed
consistently above the ``no skill'' level, although which method scored top
marks is decisively a function of flare event definition and the metric
used; there was no single winner.  Following in this paper series we ask
why the performances differ by examining implementation details 
\citep[][Paper III]{ffc3_2}, and then 
we present a novel analysis method to evaluate temporal patterns of forecasting errors
in \citep[][Paper IV]{ffc3_3}.  With these works, this
team presents a well-defined and robust methodology for evaluating solar
flare forecasting methods in both research and operational frameworks,
and today's performance benchmarks against which improvements and new
methods may be compared.
\end{abstract}


\section{Introduction}
\label{sec:intro}

Solar flares can be considered the initiating event for many Space Weather
phenomena and impacts.  The impact of solar flare radiation is almost
immediate in the case of sudden ionospheric disturbances, particularly with
{\tt M}-and {\tt X}-class flares, which disrupt radar and terrestrial
communications systems in the sunlit hemisphere.  Solar flares are also
intimately associated with other pertinent space weather phenomena such
as energetic particle storms and coronal mass ejections whose impacts
may be delayed relative to flare impacts, but can incur broader effects.
Predicting solar flare likelihood has thus long been a defined and
required operational product, now with several facilities world-wide
providing operational forecasts to a variety of customers.

Predicting solar flares is also the ultimate test of understanding
their cause, or causes.  They have long been associated with certain morphological
aspects of solar active regions such as complex structures, strong-gradient polarity
inversion lines and indications of significant energy storage in the 
magnetic field itself \citep[see {\it e.g.}, and references cited by][]{flareprediction,params,Schrijver2007}.
The only appropriate energy source is the stored free magnetic energy in 
solar active region magnetic fields,
and the only appropriate release mechanism invokes 
magnetic reconnection and reconfiguration to release that free magnetic energy.
Indeed, as discussed below and further in \citet[][Paper III]{ffc3_2}, 
quantitative ``modern'' forecasts incorporate this physical understanding as they 
often characterize coronal magnetic energy storage by proxy, with the parametrizations 
of photospheric magnetograms.
In these contexts, however, pinpointing a unique triggering mechanism has remained 
elusive.  Alternatively, solar flares may inherently be stochastic in nature 
\citep[see for example ][]{Wheatland2000,Strugarek_etal_2014,Aschwanden_etal_2016}, thus 
essentially unpredictable in a deterministic sense.  The state
of the research is presently at a point where it is still unknown
in which regime the physics operates.   While their
heliospheric and societal impacts provides motivation for predicting
these energetic events, success or failure at forecasting also provides
a key indicator as to whether stochastic physics is or is not involved.

In 2009, the first in a series of workshops was held to compare and 
evaluate the newly-emerging plethora of methods aimed at distinguishing
solar active regions with imminent flare threat.  Data from the Solar 
and Heliospheric Observatory \citep[SoHO;][]{soho} and specifically
the Michelson Doppler Imager \citep[MDI;][]{mdi} were provided to 
the methods for analysis.  The performance results \citep[see][]{allclear} 
are of secondary importance to the methodology that was established, 
identifying the importance of common definitions and standard metrics
when determining what constitutes ``good performance.''

During Solar Cycle 24, the availability of significantly
improved data sources, such as the Helioseismic and
Magnetic Imager (HMI) on the Solar Dynamics Observatory
\citep[SDO][]{sdo,hmi,hmi_cal,hmi_invert,hmi_pipe,sdo}, has made possible
a growing variety of flare forecasting systems that are running in an
operational mode (some of which were in development phase in 2009).
Consequently, an international collaboration effort was initiated
through the Center for International Collaborative Research (CICR),
at the Institute for Space-Earth Environmental Research (ISEE), Nagoya
University, Japan, to bring together the operational forecasting
teams from a variety of institutions (government, private, academic)
to evaluate the performance of different techniques.  The goals of that
workshop and subsequent analysis are to (1) establish benchmarks and
comparison methodologies for operational flare-forecasting facilities,
and (2) begin to understand what particular forecasting methodologies
enable the best forecasting performance.

The participating systems are listed in Section\,\ref{sec:methods}
with additional relevant (unpublished) details elaborated upon in
Appendix\,\ref{sec:appendix_methods}.  Although additional research
into improving forecasts is being published frequently of late
\citep{BobraCouvidat2015,Nishizuka_etal_2017,Florios_etal_2018},
for this research the comparisons were limited to those truly running
in an operational manner, which the group describes as ``providing a
forecast on a routine, consistent basis using only data available prior
to the issuance time.''  Many methods, especially the long-standing
government-institutional methods, rely on sunspot classification and
historical flaring rates \citep{McIntosh1990,flareprediction}.  A few,
now, are employing more sophisticated analysis of the host sunspot
groups and statistical classifiers or machine-learning algorithms.
Forecasts were not required to be fully automatic -- human
intervention, a ``Forecaster In The Loop''(FITL) was explicitly allowed.  Providing a forecast on a daily basis
was also not a requirement, although as an
operational system, not doing so was effectively penalized by the
evaluation metrics, as described in Section\,\ref{sec:eval}.  No further
restrictions were placed on the data employed or interval used for
training, except that it could not overlap with the testing interval
(see Section~\ref{sec:event_defs}).  The impacts of long- vs short-
training intervals ({\it e.g.} whether more than one solar cycle was
used for training the method) and other details are discussed further
in Paper III.

The participants provided forecasts for an agreed-upon interval with
agreed-upon event definitions as described in Section\,\ref{sec:event_defs}.
Representatives from most participating groups attended (in person
or remotely) a 3-day workshop during which the approaches and initial results
were discussed in depth.  The results of those days, plus further discussions 
and analysis which occurred in the subsequent months, are now presented here
and in Papers III, IV.

\section{Comparison Methodology}
\label{sec:methods}

The participating facilities and methods (with their monikers and published
references, as available) are listed in Table\,\ref{tbl:methods},
and specific details which are not available by published literature
(or modifications that have been made since the relevant publications)
are briefly described in Appendix\,\ref{sec:appendix_methods}. 
Some methods have multiple
options for producing forecasts, and those are also delineated both
in Table\,\ref{tbl:methods} and Appendix\,\ref{sec:appendix_methods}.
In Paper III  we distinguish the methods according to broad categorizations
of their implementations, such as data sources, training intervals,
imposed limits, forecast approach ({\it e.g.}, statistical, FITL) {\it etc.}, and 
hence we leave such level of detail to that paper.

\begin{longrotatetable}
\begin{deluxetable}{m{6cm}m{6cm}m{2.5cm}m{0.8cm}m{4cm}}
\tabletypesize{\scriptsize}
\tablewidth{19cm}
\tablecaption{Participating Operational Forecasting Methods (Alphabetical by Label Used) \label{tbl:methods}}
\tablehead{
\colhead{Institution} & \colhead{Name of Method/Code$^\dagger$} & \colhead{Label} & \colhead{Symbol} & \colhead{Reference(s)}}
\startdata
ESA/SSA A-EFFORT Service & Athens Effective Solar Flare Forcasting & A-EFFORT & \raisebox{-0.0cm}{\includegraphics[width=0.25cm,height=0.25cm,trim=20mm 20mm 20mm 20mm angle=0]{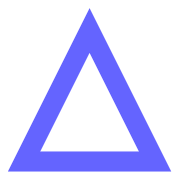}} &  \citet{GeorgoulisRust2007}\\ \hline
Korean Meteorological Administration \& Kyung Hee University & Automatic McIntosh-based Occurrence probability of Solar activity & AMOS & \raisebox{-0.0cm}{\includegraphics[width=0.25cm,height=0.25cm,trim=20mm 20mm 20mm 20mm angle=0]{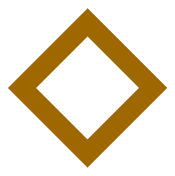}} & \citet{amos}\\ \hline
University of Bradford (UK) & Automated Solar Activity Prediction & ASAP & \raisebox{-0.0cm}{\includegraphics[width=0.25cm,height=0.25cm,trim=20mm 20mm 20mm 20mm angle=0]{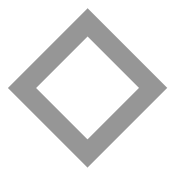}} &  \citet{ColakQahwaji2008,ColakQahwaji2009} \\ \hline
Korean Space Weather Center (by SELab, Inc) & Automatic Solar Synoptic Analyzer & ASSA &\raisebox{-0.0cm}{\includegraphics[width=0.25cm,height=0.25cm,trim=20mm 20mm 20mm 20mm angle=0]{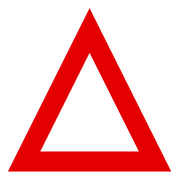}}  & \citet{ASSA}, \citet{ASSA_DOC} \\ \hline
Bureau of Meteorology (Australia) & FlarecastII & BOM  &\raisebox{-0.0cm}{\includegraphics[width=0.25cm,height=0.25cm,trim=20mm 20mm 20mm 20mm angle=0]{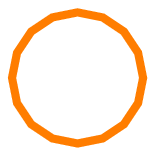}}  & \citet{Steward_etal_2011,Steward_etal_2017}\\ \hline
120-day No-Skill Forecast & Constructed from NOAA event lists & CLIM120  & \raisebox{-0.0cm}{\includegraphics[width=0.25cm,height=0.25cm,trim=20mm 20mm 20mm 20mm angle=0]{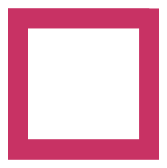}} & \citet{SharpeMurray2017} \\ \hline
NorthWest Research Associates (US) & Discriminant Analysis Flare Forecasting System & DAFFS  &\raisebox{-0.0cm}{\includegraphics[width=0.25cm,height=0.25cm,trim=20mm 20mm 20mm 20mm angle=0]{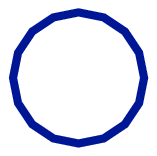}}  & \citet{nci_daffs} \\ \hline
 '' '' & GONG+GOES only & DAFFS-G  &\raisebox{-0.0cm}{\includegraphics[width=0.25cm,height=0.25cm,trim=20mm 20mm 20mm 20mm angle=0]{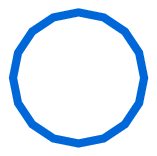}}  &  '' ''  \\ \hline
NASA/Marshall Space Flight Center (US) & MAG4 (+according to & MAG4W & \raisebox{-0.0cm}{\includegraphics[width=0.25cm,height=0.25cm,trim=20mm 20mm 20mm 20mm angle=0]{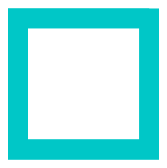}} & \citet{Falconer_etal_2011}; \\ 
'' '' & magnetogram source  & MAG4WF & \raisebox{-0.00cm}{\includegraphics[width=0.25cm,height=0.25cm,trim=20mm 20mm 20mm 20mm angle=0]{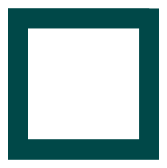}} & also see Appendix\,\ref{sec:appendix_methods} \\ 
'' '' & and flare-history  & MAG4VW & \raisebox{0.00cm}{\includegraphics[width=0.25cm,height=0.25cm,trim=20mm 20mm 20mm 20mm angle=0]{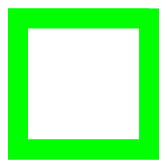}} &  \\ 
'' '' & inclusion) & MAG4VWF & \raisebox{-0.00cm}{\includegraphics[width=0.25cm,height=0.25cm,trim=20mm 20mm 20mm 20mm angle=0]{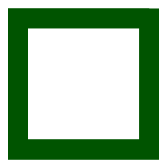}} &  \\ \hline
Trinity College Dublin (Ireland) & SolarMonitor.org Flare Prediction System (FPS)  & MCSTAT  & \raisebox{-0.0cm}{\includegraphics[width=0.25cm,height=0.25cm,trim=20mm 20mm 20mm 20mm angle=0]{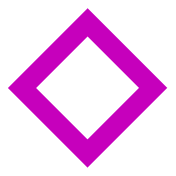}} & \citet{gallagheretal02,Bloomfield_etal_2012}\\ \hline
 '' '' & FPS with evolutionary history & MCEVOL  & \raisebox{-0.0cm}{\includegraphics[width=0.25cm,height=0.25cm,trim=20mm 20mm 20mm 20mm angle=0]{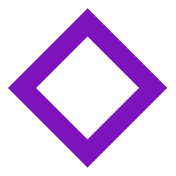}} & \citet{McCloskey_etal_2018} \\ \hline
MetOffice (UK)  &  Met Office Space Weather Operational Center human-edited forecasts & MOSWOC  & \raisebox{-0.0cm}{\includegraphics[width=0.25cm,height=0.25cm,trim=20mm 20mm 20mm 20mm angle=0]{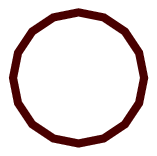}}  & \citet{Murray_etal_2017}\\ \hline
National Institute of Information and Communications Technology (Japan) &  NICT-human  & NICT & \raisebox{-0.0cm}{\includegraphics[width=0.25cm,height=0.25cm,trim=20mm 20mm 20mm 20mm angle=0]{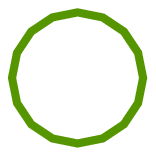}} &  \citet{Kubo_Den_Ishii_2017}\\ \hline
New Jersey Institute of Technology (UK) & NJIT-helicity & NJIT  & \raisebox{-0.0cm}{\includegraphics[width=0.25cm,height=0.25cm,trim=20mm 20mm 20mm 20mm angle=0]{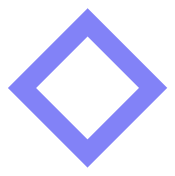}} & \citet{Park_Chae_Wang_2010}\\ \hline
NOAA/Space Weather Prediction Center (US) & & NOAA & \raisebox{-0.0cm}{\includegraphics[width=0.25cm,height=0.25cm,trim=20mm 20mm 20mm 20mm angle=0]{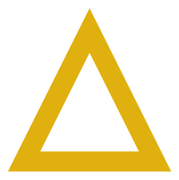}}  & \citet{Crown2012}\\ \hline
Royal Observatory Belgium Regional Warning Center & Solar Influences Data Analysis Center human-generated & SIDC & \raisebox{-0.0cm}{\includegraphics[width=0.25cm,height=0.25cm,trim=20mm 20mm 20mm 20mm angle=0]{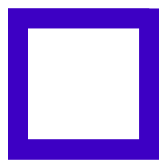}}  & \citet{Berghmans_etal_2005,Devos_etal_2014}\\ \hline
\enddata
\flushleft{$\dagger$: if applicable}
\end{deluxetable}
\end{longrotatetable}
\clearpage

\subsection{Event Definitions and Testing Interval}
\label{sec:event_defs}

The participants agreed on a testing interval of 01 January 2016 --
31 December 2017 for evaluating forecasts.  This is arguably a very
short testing interval; in the present situation, it was chosen to
balance both training and testing data for those methods relying upon
data from {\it SDO}/HMI, since the near-real-time data from HMI are 
only available from late 2012.  The resulting activity levels are summarized in
Table\,\ref{tbl:eventrates}.  Evaluation was performed on full-disk
forecasts only, to avoid the requirement of standardizing the different
active-region identification methods in use (combining region-based
forecasts to full disk is described in Appendix\,\ref{sec:appendix_FD}).

Event definition choices were dictated by the need for common definitions
across methods and the fact that these are operational methods,
hence most already produce forecasts that match the
NOAA/SWPC-established event definition and timings.

\begin{table}[t]
\caption{24\,hr Event Rates for 2016.01.01 -- 2017.12.31}
\label{tbl:eventrates}
\begin{center}
\begin{tabular}{|lccc|} \hline
Class & \# Quiet Days & \# Event Days & Climatology (Event Day Rate)\\ \hline
\Cp\  & 543 & 188  & 0.257\\
\Mp\  & 705 & 26 & 0.036\\
\Xp\ & 728 & 3 & 0.004\\ \hline
\end{tabular}
\end{center}
\end{table}

As such, the group agreed upon event thresholds as ``lower-limits
plus exceedance'' following the NOAA/SWPC definition, based on the
NOAA {\it Geostationary Observing Earth Satellite} (GOES) 
X-Ray Sensor (XRS) 1\,--\,8\,\AA\ bands: \Cp\ and \Mp\ corresponding to lower limits of
$1.0\times10^{-6}$ and $1.0\times10^{-5}\,{\rm W m}^{-2}$, respectively, with
no upper limit ({\it i.e.},``exceedance'' forecasts).  All forecasts were put onto an 
exceedance basis; calculating 
exceedance forecasts from category-limited forecasts ({\it i.e.} including an upper limit), as were provided
by some methods, is discussed in Appendix\,\ref{sec:appendix_combos}.
No background or pre-flare subtraction was performed for the 
evaluation data, which is consistent with none generally being performed
by any operational method during either training or event prediction 
\citep[see also][for a discussion on the impact of background subtraction.]{Wheatland2005}.
The event definitions include 24\,hr validity periods and effectively 
$0$\,hr latencies (the time periods between forecast issuance and the start of the
validity period) for the initial comparisons ({\it i.e.} only ``one-day''
forecasts, not longer-range forecasts).  Longer effective latencies may be
implied due to data acquisition times, but these are ignored here 
for delays $<1$\,hr.
Additionally, it is noted that a number of centers produce additional
forecasts (with variations in frequency of forecast, event thresholds,
latencies, or validity periods); for this comparison, we chose the event
definitions to assure the most overlap between methods.  We refer 
now to these two event definitions using the shorthand ``\CC''
and ``\MM'', noting that the nomenclature includes all three
aspects of the event definition (thresholds, latency in hours, and validity period in hours).

The \CC\ exceedance definition provided 188 event-days, and
the \MM\ exceedance definition provided 26 event-days over
the 731 days of the testing interval (2016 was a leap-year; see
Table\,\ref{tbl:eventrates}).  Not all methods produce \CC\
forecasts.  While most methods produce a forecast for \Xp\ 
($1.0\times10^{-4}\,{\rm W m}^{-2}$ and larger), in practice the short
testing interval produced too few of these largest events to provide
meaningful evaluations.

Most methods issue a forecast in the neighborhood of midnight Universal
Time.  Within approximately one hour, any discrepancy from midnight
was ignored.  Beyond that, the discrepancies in event lists would become
problematic.  For methods which issue forecasts significantly different
from midnight (SIDC at 12:30\,UT, NICT at 06:00\,UT), custom event lists were
constructed based on that issuance time.  Although these custom lists do change
the number of events slightly (\CC\ becomes 183 and 185 event-days for
NICT and SIDC respectively; \MM\ becomes 27 event-days for both),
they provide the most appropriate approach to enable cross-comparisons.
Almost all methods issue multiple forecasts throughout the
day; in the course of these comparisons the forecast issued closest to 
midnight Universal Time (UT) was used and others were ignored.  

\subsection{Standard Metrics and Evaluation Tools}
\label{sec:eval}

Different performance metrics inform on different performance
aspects.  This is discussed in \citet{JolliffeStephenson2012}
and other references specifically with regards to flare forecasting in
\citet{Bloomfield_etal_2012,allclear,Kubo_Den_Ishii_2017,Steward_etal_2017,Murray_etal_2018}.
Hence, we present a number of metrics and evaluation tools,
but for brevity we refer to any of the above references
for the definitions of specific metrics\footnote{See also {\tt
http://www.cawcr.gov.au/projects/verification/\#What\_makes\_a\_forecast\_good}
and {\tt
https://www.nssl.noaa.gov/users/brooks/public\_html/feda/note/reliroc.html}
for broad discussion and numerous definitions}.

Graphical representations of performance are used due to the wealth of
information available in a compact form.  Reliability Plots (also known as
Attribute Diagrams) plot bins of the predicted probability against
the observed number of instances in that event frequency bin.  A perfect
reliability displays points along the $x=y$ line.  A perfect
forecast is one in which an event is only and always predicted with a
probability of 100\%; such a service will
only have points in the first and last probability bins.
Also included in these plots are the climatological rate (event rate)
for the testing period (a $y$=constant line at the event rate for that
testing period) and the ``no skill'' line which is defined as the 
bisector between the testing-interval climatology and the ``perfectly reliable'' $x=y$ line.
Additionally, we indicate the relative population of the full sample proportion
of forecasts within each bin.

Relative (Receiver) Operating Characteristic (Curve) or ``ROC'' diagrams
are constructed by plotting the Probability of Detection (POD) {\it vs.} the
Probability of False Detection (POFD) as a threshold is varied by which 
a forecast outcome becomes a ``yes'' forecast.  This threshold is 
commonly referred to as the Probability Threshold $P_{\rm th}$ as it 
is applied to forecast probabilities, but is applied here even though
some methods may not strictly produce probabilities.
ROC diagrams measure resolution but not reliability.  ROC diagrams include the
$x=y$ line to indicate ``no skill''; on a ROC plot, perfect forecasts
trace the path from $(0,0)$ to $(0,1)$ to $(1,1)$.

Supplementing the graphical evaluation tools are quantitative metrics.
Skill score metrics in particular compare performance to that of a
reference forecast.  These are normalized such that perfect forecasts
result in a metric of 1.0, and ``no skill'' as compared to the reference
results in 0.0.  The reference forecast may take various forms; commonly
used is the climatology of the testing period or a random forecast
\citep{JolliffeStephenson2012}, but it may be any other valid forecast
method.

The Reliability Plots can be summarized by the Brier Skill Score (BSS),
a metric based on the probability forecasts, and for which the reference
is specifically the no-skill climatological forecast of the testing period
(see Table\,\ref{tbl:eventrates}).  This metric answers the question,
``how well did this method do compared to the underlying climatology?''.

The ROC curves are summarized here by the ROC Skill Score (ROCSS) also known
as the Gini Coefficient, both of which are related to the Area Under the Curve (AUC)
but provide more discrimination
\citep{JolliffeStephenson2012,nci_daffs}.  
The ROCSS and Gini coefficient 
are normalized such that no skill provides a score of $0.0$, and perfect 
forecasts provide a score of $1.0$.

Deterministic (or categorical) forecasts can be valuable when preparing
forecasts for a particular customer who may require a specified acceptable
rate of false alarms, for example, rather than simply a probabilistic
forecast.  Four additional metrics based on dichotomous (yes/no) forecasts
are included: the Appleman Skill Score (ApSS) uses the testing
interval to construct an ``across the board'' climatology reference
forecast (a single reference forecast according to the event day rate
in the testing interval), the Equitable Threat Score (ETS) invokes a random
forecast, and the Hanssen \& Kuiper Skill Score / Peirce Skill
Score/ True Skill Statistic (here just PSS/TSS) is the difference
between the POD and the POFD (see definitions and discussions in
\citet{Woodcock1976,Murphy1996,BarnesLeka2008,Bloomfield_etal_2012,allclear,Murray_etal_2017,
Kubo_Den_Ishii_2017}).  These metrics are all based on permutations of
the ``truth table'' entries that compare Predicted {\it vs.} Observed
outcomes, and are discussed at length in the references cited above.
Additional numeric metrics such as the Proportion Correct (PC, also called
Rate Correct or Accuracy) and the Frequency Bias (FB) \citep{JolliffeStephenson2012}
do not compare to reference forecasts per se, and may or may not have
a similar normalization as required for a true skill score.  The PC
metric is common (but can be misleadingly high even for unskilled forecasts 
in highly unbalanced samples) and the FB indicates systematic 
over- or under-forecasting, a necessary complement to the TSS metric.

A deterministic forecast is produced by imposing a threshold
$P_{\rm th}$ for assigning the probabilities or forecast outcomes to yes/no 
forecasts.   This threshold reflects a probability level for an event
at which a ``real-world'' action/no-action decision has to be taken
based on, for example, economic losses incurred from one or the other type of 
error.  This threshold is then also used for the dichotomous-based metrics 
(PC, ApSS, ETS, PSS/TSS, FB) by which that method is evaluated.
The performance of a method according to a dichotomous-based
metric may vary as a function of $P_{\rm th}$ -- this is demonstrated
in ROC curves where the vertical distance of each point of the curve 
from the no-skill $x=y$ line reflects the PSS/TSS and thus
the method's discrimination between events and non-events as $P_{\rm
th}$ is varied \citep[see the discussion in][]{allclear}.  Generally 
speaking, the methods here are either not
explicitly optimized for a particular $P_{\rm th}$ during their training
or the training method implicitly maximizes a particular metric that
effectively optimizes the system at $P_{\rm th}=0.5$. 
All but one method produced probabilistic forecasts; for the one that
did not, outputs of $0.0$ and $1.0$ were assigned ``no'' and ``yes''
forecasts, respectively.  

Hence, we adopt $P_{\rm th}=0.5$ to compute dichotomous-based metrics
for all methods.  A few methods provide custom forecasts to customers with
different $P_{\rm th}$, or routinely provide their alerts above a
particular $P_{\rm th}$, and those were invited for evaluation with a
custom $P_{\rm th}$ (none were submitted).  Unless specified
otherwise, selecting $P_{\rm th}=0.5$ for categorical-based metrics is
an allowable choice for all methods.  All probabilities for all forecast
methods accompany this publication\footnote{Leka and Park 2019, Harvard Dataverse, doi:10.7910/DVN/HYP74O}
and are thus available for readers to calculate additional metrics,
for example with $P_{\rm th}\neq 0.5$.

For all methods, missing forecasts were assigned
a probability $p=0.0$ for that day.  This is appropriate for operational
forecasts, where missed or skipped forecasts should be penalized.
Most operational methods have built in backup sources of data, forecasts,
or the ability to forecast prior climatology in the event of, for example,
data interruption (see additional details in Paper III).

We do not present the popular ``maximum TSS'' (TSS$_{\rm max}$) for 
two reasons.  First, an
``optimal $P_{\rm th}$'' with which TSS$_{\rm max}$ is calculated 
should be established based on information obtainable only
from the training interval, rather than the testing interval itself,
as is common practice.  No method supplied such a customized
$P_{\rm th}$ to use.  Determining an ``optimal $P_{\rm th}$'' from which
to achieve a maximum TSS score based on testing-period information is
not consistent with a purely operational approach.  The optimal $P_{\rm
th}$ can have a correspondence to the underlying event rate
\citep{Bloomfield_etal_2012,allclear}, which varies according to 
the solar cycle and from one cycle to the next as discussed 
below.\footnote{Some methods ({\it e.g.} A-EFFORT) do establish 
optimal $P_{\rm th}$ levels during training and apply them in order to issue alerts.
They elected to not invoke these $P_{\rm th}$ for the evaluations here.}
Hence, there is limited ``actionable information'' in
determining the optimal $P_{\rm th}$ from a training period for future forecasting.  
Second, the $P_{\rm th}$ for each method used to achieve TSS$_{\rm max}$
will differ from each other and will depend on the event definition,
so interpreting these results is challenging \citep[see discussion
in][]{allclear}.  That being said, one can roughly estimate TSS$_{\rm
max}$ for each method from the shape of its ROC plot ({\it i.e.} the point of maximum
vertical departure from the no-skill $x=y$ line).

\begin{figure}[t]
\centerline{\includegraphics[width=0.95\textwidth,clip, trim = 10mm 0mm 0mm 5mm, angle=0]{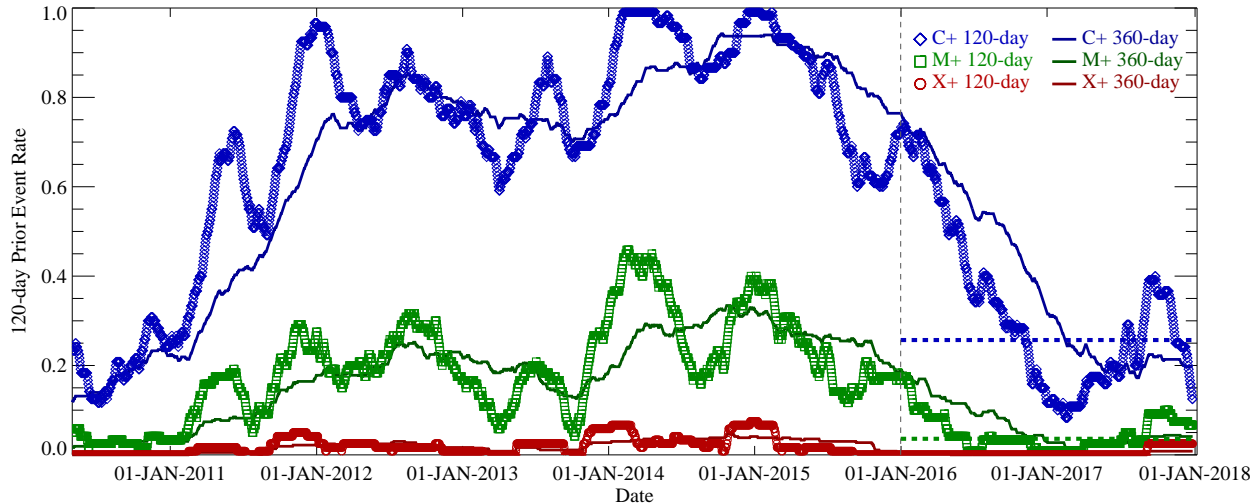}}
\caption{The ``120-day prior climatology'' and ``360-day prior climatology are 
plotted for the  \CC\ and \MM\ event
definitions, plus the same for an \Xp\ threshold for completeness, from
the start of the SDO mission (2010.05.01) through the testing interval,
whose start is indicated by a vertical dashed line.  The climatological
event rate of the testing interval is indicated by horizontal dashed lines
over that time period.  Each symbol-point (as indicated) represents
the daily full-disk event rate for the prior 120 days (up until but not
including the date on which the point falls), similarly for the 
curves indicating the 360-day prior climatology.  The 120-day prior climatology
is used as the unskilled reference forecast in the ``MSESS\_clim'' and
``ApSS\_clim'' metrics in Figure\,\ref{fig:stats}.}
\label{fig:clims}
\end{figure}

\subsection{Highlighted Metrics: Comparison against No-Skill Operational Forecasts}

All metrics discussed thus far explicitly evaluate the performance
of forecasts against the outcome of the testing interval.  In true
operational settings, however, an appropriate reference forecast
against which to judge performance is more appropriately the best  ``unskilled''
forecast available \citep{SharpeMurray2017,Murray_etal_2017}.  In other
words, for operational forecasting it is appropriate to separately and
specifically ask ``to what extent is the method in question an improvement
beyond what would be otherwise available by simply using an unskilled forecast?''
If a forecasting method cannot perform better than this unskilled forecast,
then it does not add any skill or value beyond that unskilled forecast.

To construct a ``no-skill'' forecast for day $t$ for the event
definition in question, we use an event rate determined over the prior
$N$ days up to and including $t-1$.  The resulting event rate is then used as the
reference forecast's predicted probability for that date $t$.  We choose
$N=120$ days as suggested by \citet{SharpeMurray2017}.  This unskilled
reference forecast does vary, as shown in Figure\,\ref{fig:clims} --
in particular decreasing from $>0.5$ to $<0.5$ for \CC\ within
the testing interval.  Its abrupt variation on short timescales ({\it e.g.}
around September 2017, see
also \citet{SharpeMurray2017} Figure 5) likely reflects active-region
recurrence patterns and space weather effects rather than reflecting
longer-range climatology \citep[see discussion on climatology variations
in][and the $360$-day prior climatology curves also shown here in 
Figure\,\ref{fig:clims}]{McCloskey_etal_2018}.  However, a $120$-day prior climatology forecast
avoids significant lag 
against the fairly rapid event-rate changes that
occur at the beginning and end of the solar magnetic cycles 
evident in the $360$-day prior climatology curves.  Either  provides
a valid unskilled forecast and a valid reference forecast for associated metrics, 
with expected performance differences and resulting scores -- as would a ``no-skill''
forecast using yet another value for $N$.  The 120-day prior climatology forecast (``CLIM120'') 
is included for evaluation along with all other methods as a ``sanity check''
on the performance of this reference forecast.


Two metrics are constructed using this unskilled forecast as the
reference.  A metric ``MSESS\_clim'' is analogous to the Brier Skill
score as based on the mean square error (MSE) of the forecast probabilities.
However, instead of the testing-period climatology as defined for the BSS,
the MSESS\_clim uses the prior 120-day event rate (``120-day prior
climatology'') as the reference forecast.  Analogously, we compute an
Appleman skill score for which the ``across-the-board'' forecast for
any given day is dictated by this reference; the resulting accuracy is
computed and used as the reference forecast in the ``ApSS\_clim'' score.

\section{The Method Performances}
\label{sec:perf}

\begin{figure}
\centerline{\includegraphics[width=0.68\textwidth,clip, trim = 0mm 0mm 0mm 0mm, angle=0]{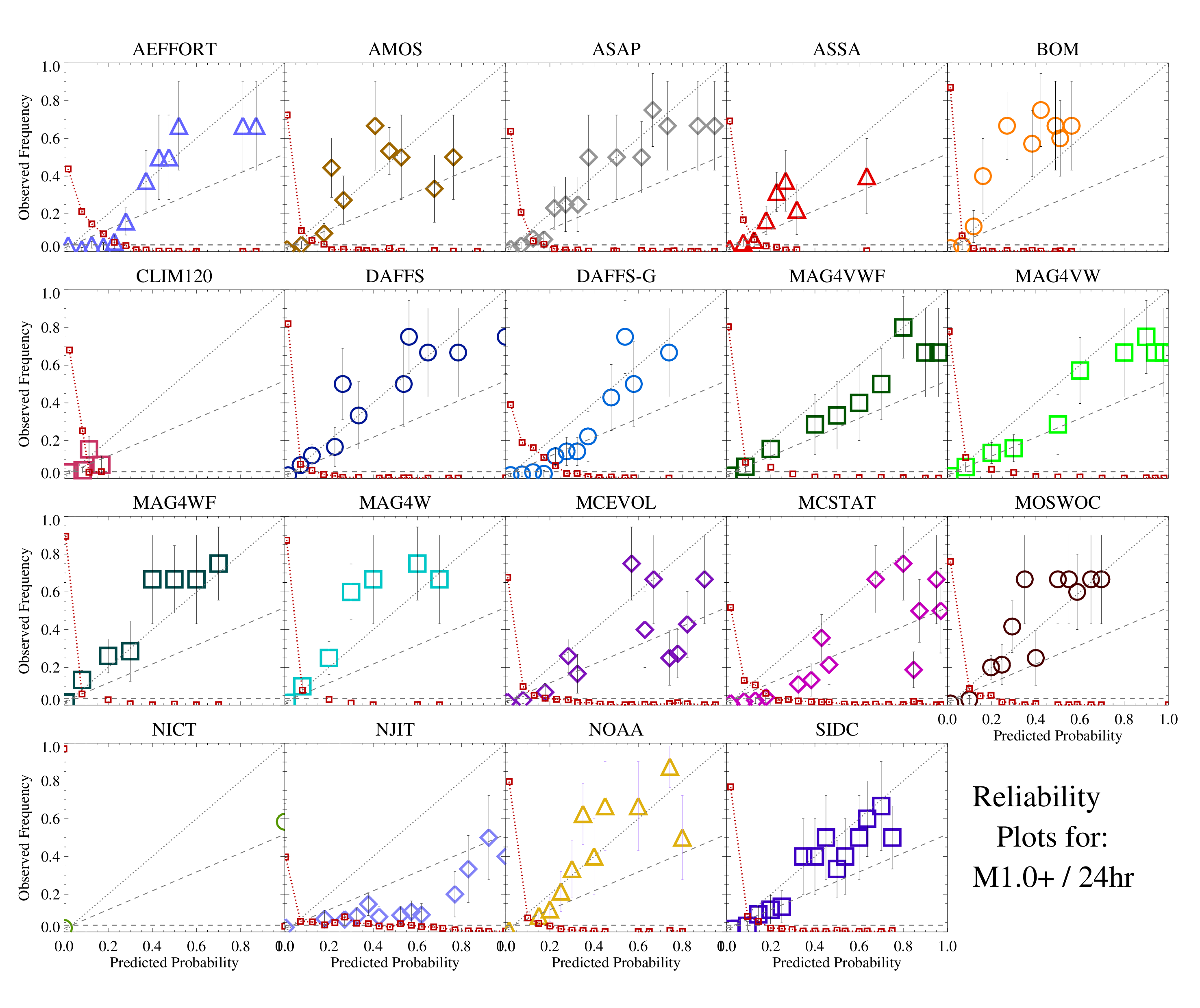}}
\vspace{-0.3cm}
\centerline{\includegraphics[width=0.68\textwidth,clip, trim = 0mm 0mm 0mm 0mm, angle=0]{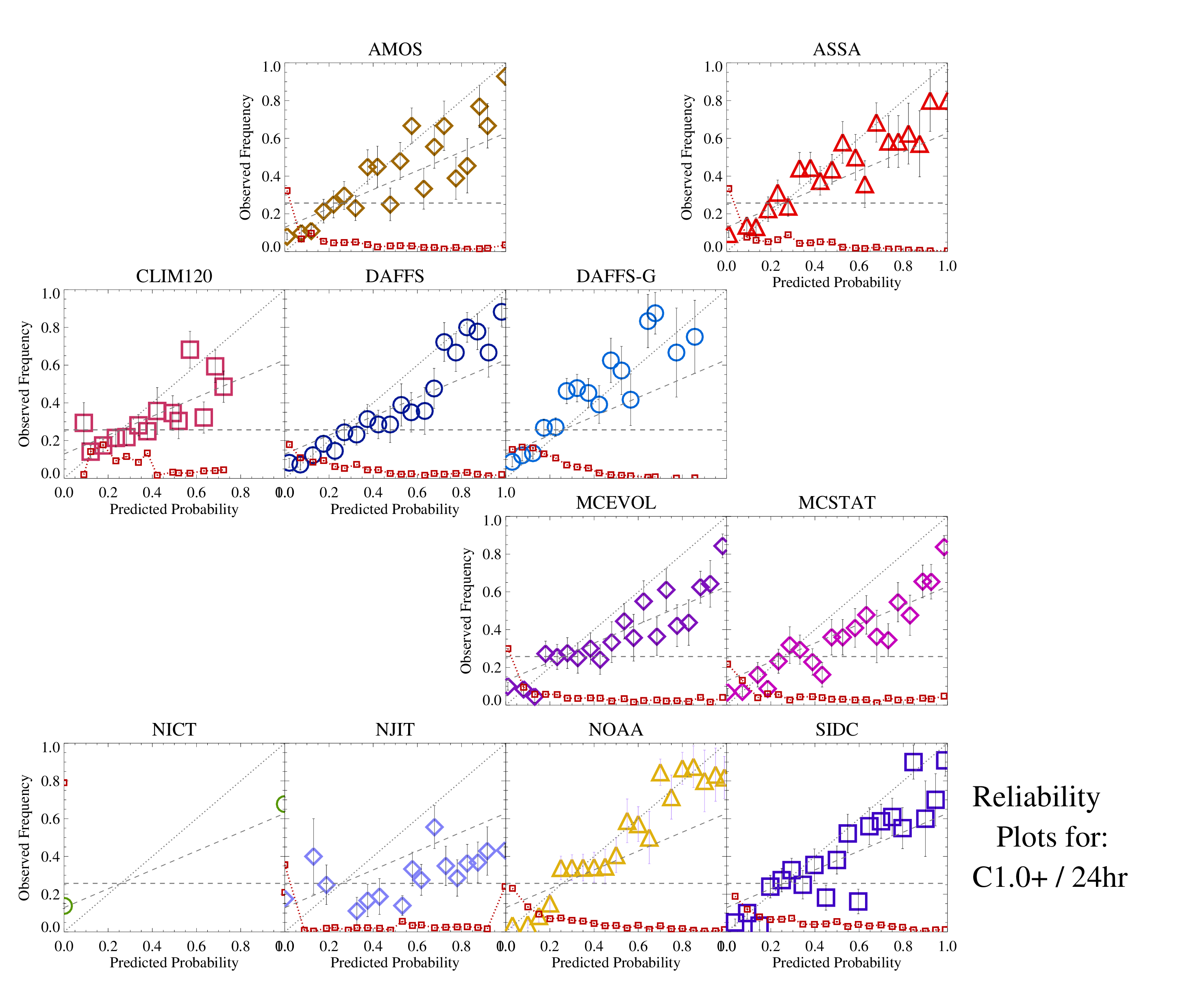}}
\vspace{-0.6cm}
\caption{Reliability plots (Attribute Diagrams) for each method,
indicating the performance of the probabilistic forecasts as named, the
$x=y$ ``perfect reliability'' line, the (horizontal) climatology level
(dashed line), and the ``no skill'' line (dotted line) that lies between
the two.  Additionally (red dotted line and small square) is 
the fraction of the total sample
for which a forecast exists for each bin.  Each method has an assigned
color / symbol combination (Table\,\ref{tbl:methods}), where related
methods ({\it e.g.} from the same institution) have the same symbols
and are plotted with colors in the same family (``nearby'' in hue).
Results are shown for \MM\ (top) and \CC\ (bottom); fewer methods predict
the latter than the former.  Results were not calculated for \Xp\ due
to extremely small number of events in the testing interval.}
\label{fig:reliability}
\end{figure}

\begin{figure}
\centerline{\includegraphics[width=0.68\textwidth,clip, trim = 0mm 0mm 0mm 0mm, angle=0]{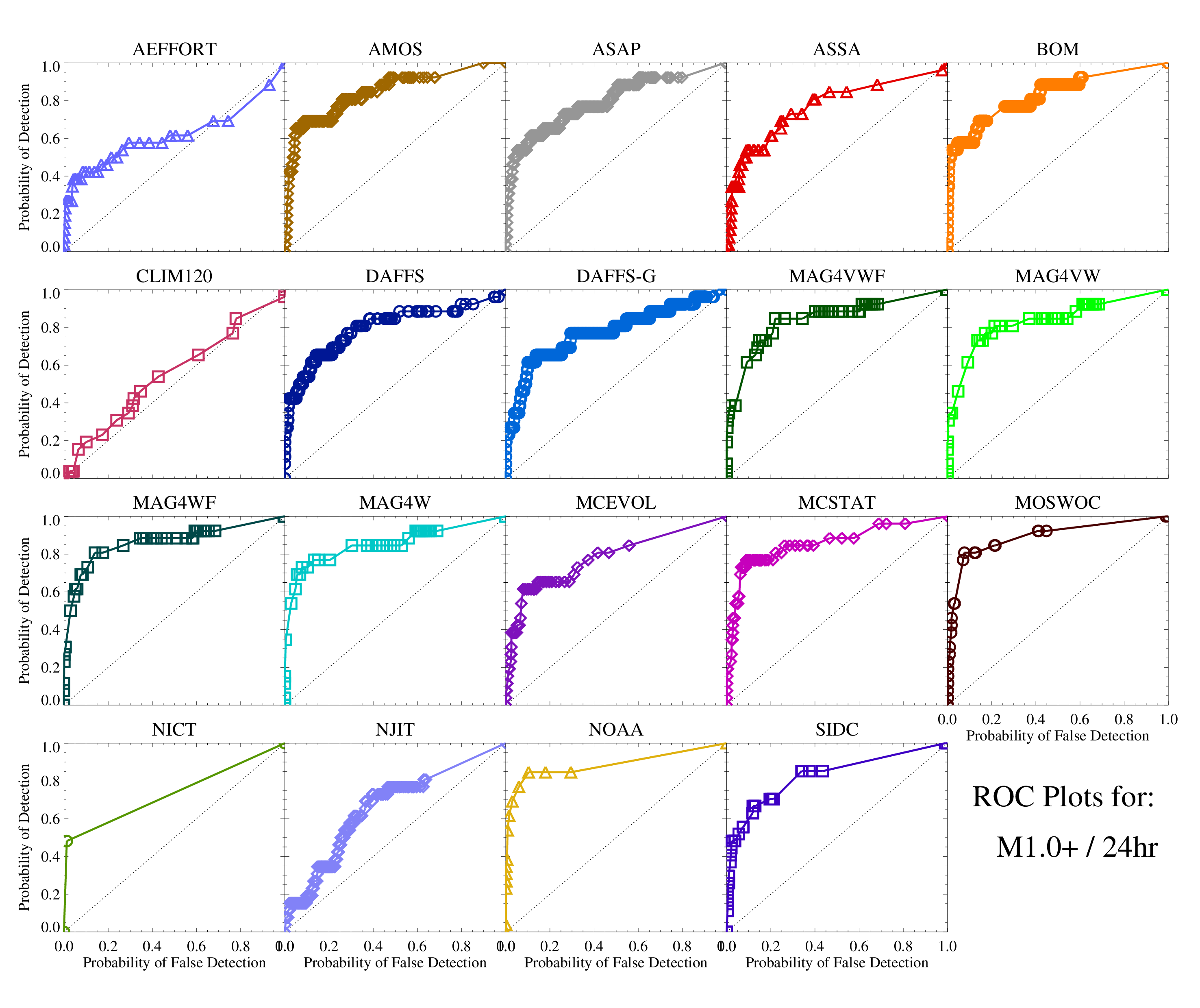}}
\centerline{\includegraphics[width=0.68\textwidth,clip, trim = 0mm 0mm 0mm 0mm, angle=0]{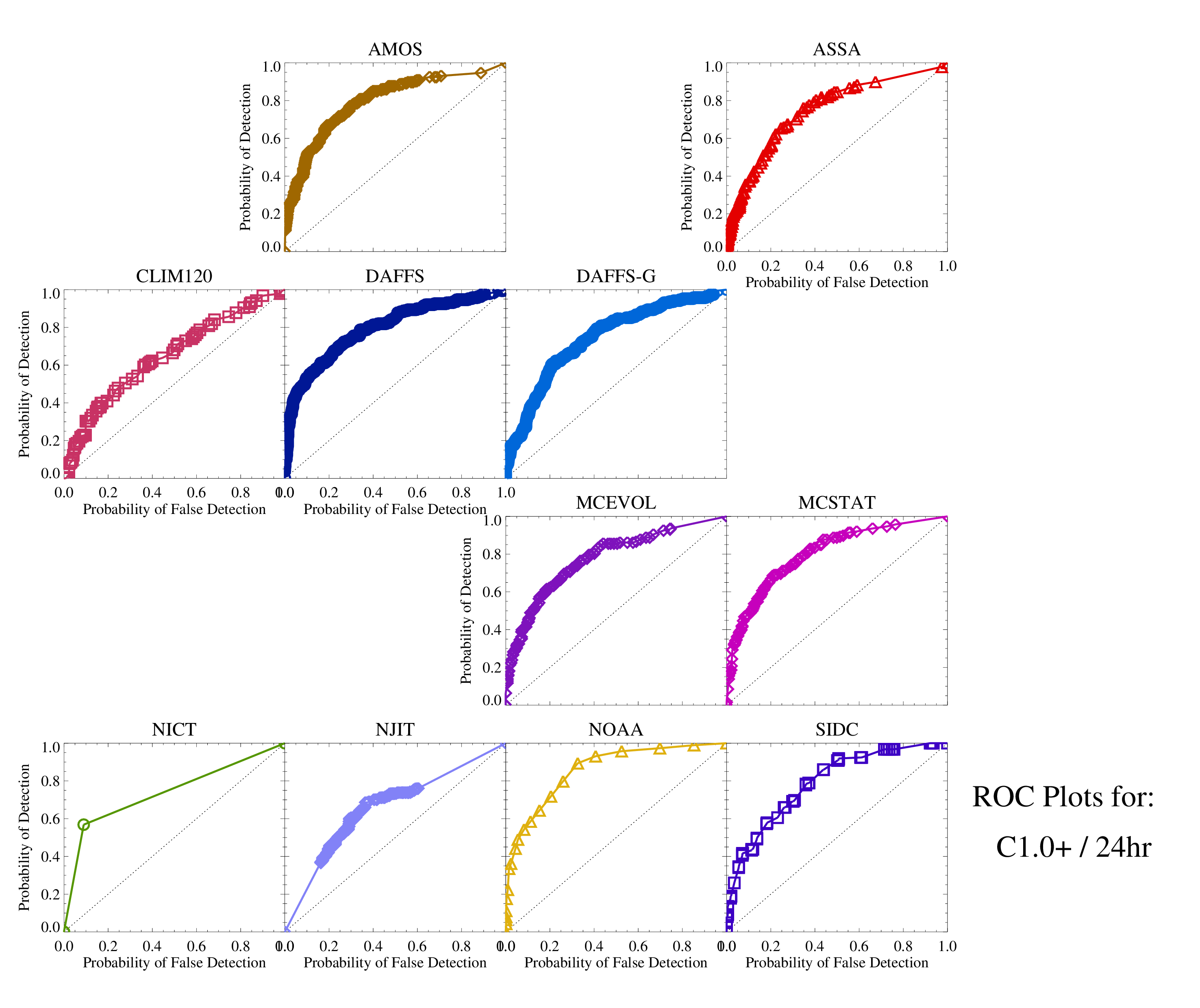}}
\caption{Relative Operating Characteristic plots with the $x=y$ ``no skill'' line, following the 
color/symbol scheme of Figure\,\ref{fig:reliability}.}
\label{fig:rocs}
\end{figure}

\begin{figure}
\centerline{\includegraphics[width=0.97\textwidth,clip, trim = 0mm 0mm 0mm 0mm, angle=0]{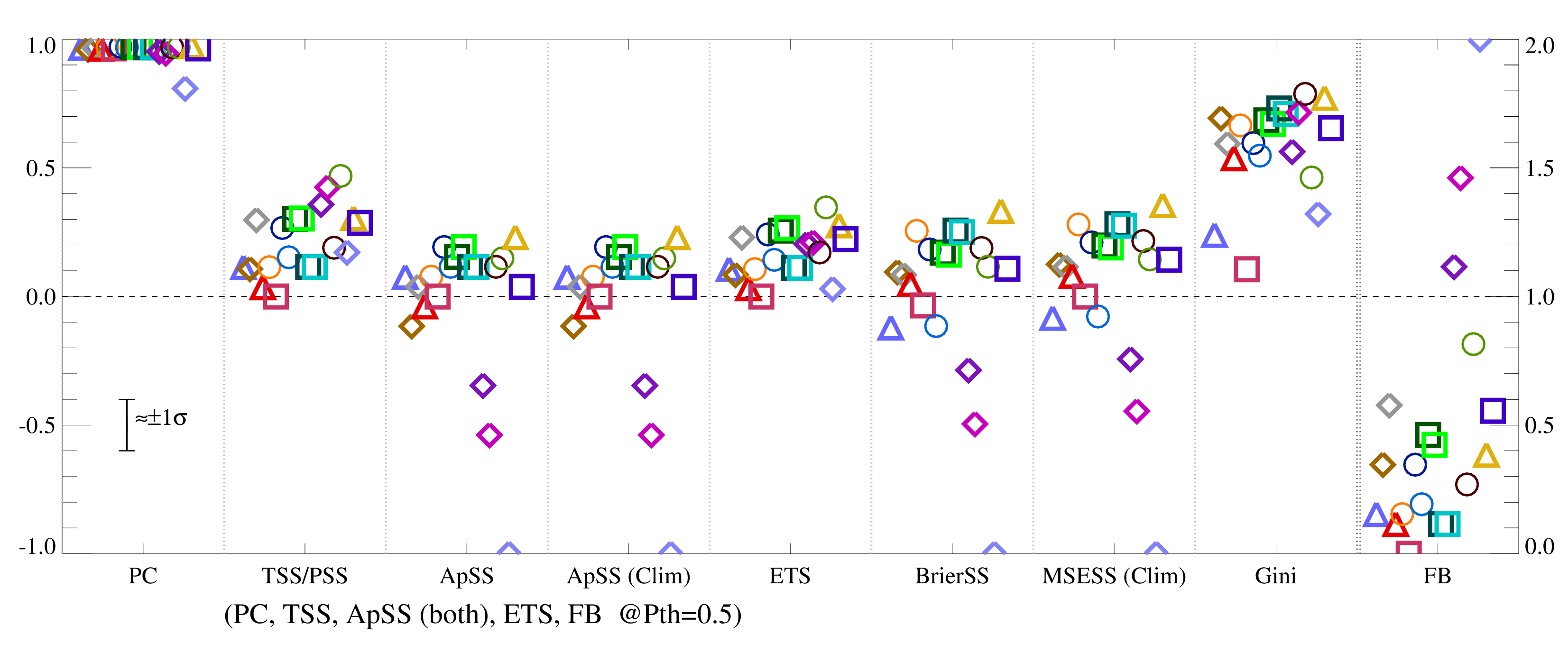}}
\centerline{\includegraphics[width=0.97\textwidth,clip, trim = 0mm 0mm 0mm 0mm, angle=0]{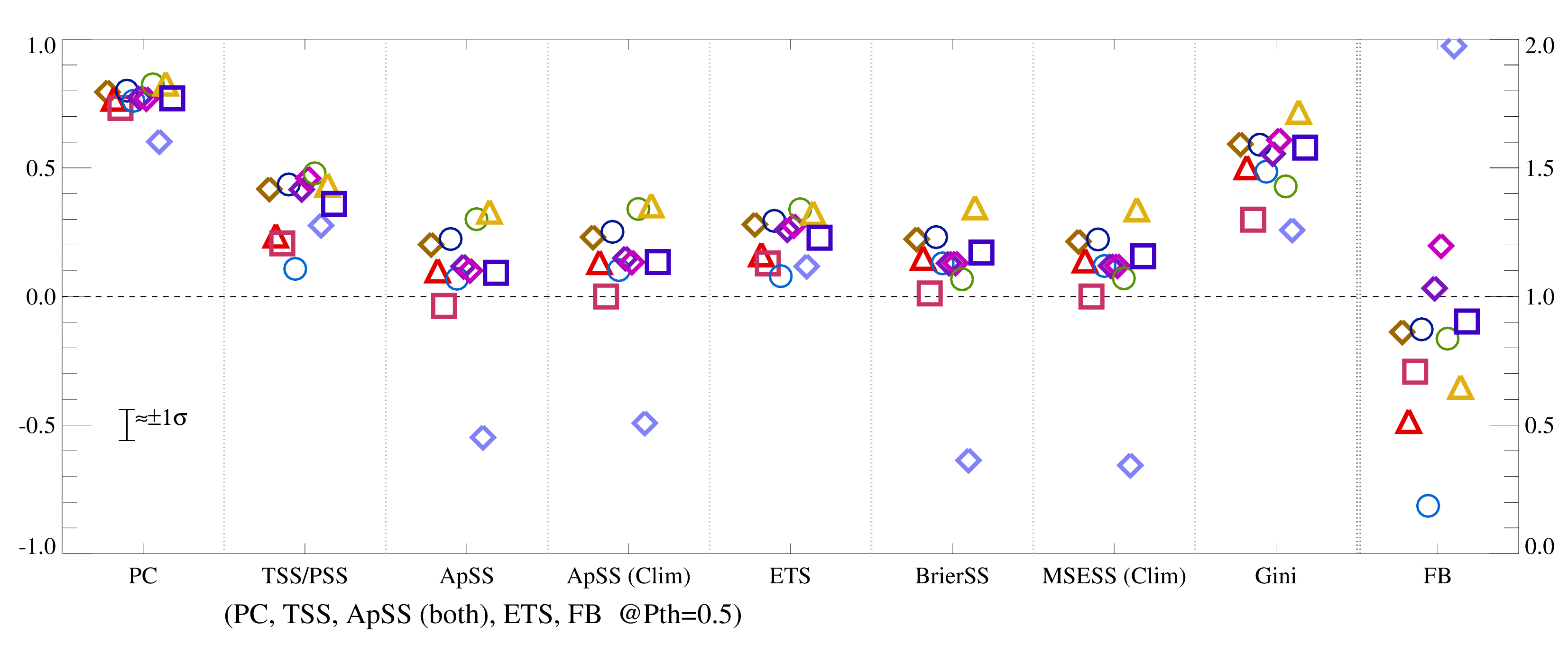}}
\caption{Results from the direct comparison of flare forecasting methods
for a variety of performance metrics (Left to Right): the Proportion
Correct, the True Skill Statistic / Peirce Skill Score, Appleman Skill
Score (testing period), the Appleman Skill Score with the 120-day
prior climatology reference forecast, the Equitable Threat Score,
Brier Skill Score, a Mean Square Error Skill Score with the 120-day
prior climatology as a reference forecast, and the Gini Coefficient.
A lower limit of $-1.0$ was imposed for the plotting.  The final
metric is the Frequency Bias, whose displayed range is indicated on
the right-hand axis; a $+2.0$ limit was placed on this plot.  Metrics
based on ``truth tables'' are calculated using $P_{\rm th}=0.5$;
the Brier Skill Score and Mean Square Error Skill Score (clim),
and Gini Coefficient are independent of $P_{\rm th}$.  The symbols
follow the scheme in Figures\,\ref{fig:reliability},\,\ref{fig:rocs}
and are offset slightly in the $x$-dir for clarity in the same order
as they appear in Figures\,\ref{fig:reliability},\,\ref{fig:rocs}, and
Table\,\ref{tbl:methods}).  Results are shown for \MM\ (top) and \CC\
(bottom); fewer methods predict the latter than the former.  Results were
not calculated for \Xp\ due to the extremely small number of events in
the testing interval.}
\label{fig:stats}
\end{figure}

Results are shown here for the metrics and evaluation methodology
described in the previous section.  Of note, if a particular method is
highlighted in the text as an example of a particular trend it will
rarely be the only example, and such a call-out does not mean other
methods are exempt from said trend.  Such call-outs refer to \MM\
results unless otherwise noted.

First, in Figure\,\ref{fig:reliability} the Reliability Diagrams
(Attribute Diagrams) are shown, comparing predicted probabilities
to the observed frequencies across 20 probability bins.  The predicted
probabilities are 
indicated on the $x$-axis by the average of the probabilities in that bin.
Points in each bin are accumulated, and thus accurately
reflect the distribution whether from continuous probabilities or 
discrete forecast probabilities.
This figure also displays the symbol and color schemes devised to both
compare methods and inter-compare between related methods ({\it e.g.}
variations from the same institution, {\it c.f.} Table\,\ref{tbl:methods}).
Most methods provided a form of \MM\ forecasts (natively, or computed
as per Appendix\,\ref{sec:appendix_combos}).  A subset of methods
also produce forecasts for \CC\, and those are displayed as well.
The decision regarding whether to produce forecasts for these smaller flares
rests on the facility or agency according to resources, customer needs, and
perceived threat; if publicly available, these forecasts were included.
Most methods do provide a forecast for \Xp, however the number of
events was so small during the testing period as to be uninformative
(see Table\,\ref{tbl:eventrates}).   The error bars are determined by the
number of points and events in each bin \citep{Wheatland2005,allclear}:
for a reliability value $R$ in a particular bin, $\sigma_{\rm R}=(R ((1-R)/N_{\rm
bin}+3))^{1/2}$ with $N_{\rm bin}$ being the number of points in that bin.

The Reliability Diagrams graphically display trends of over-forecasting
({\it cf.} MCSTAT) or under-forecasting ({\it cf.} MAG4W) the issued
probabilities.  Some methods more systematically perform errors of
one type ({\it e.g.} BOM) while others display a mix according to the
probability bin ({\it e.g.} AMOS) but not an obvious 
dominance of one
error or the other.  The reliability plots also highlight that some
probabilistic methods provide predictions covering the full range of
probabilities ({\it e.g.} MAG4VWF) while others do not provide predictions
at the highest probabilities ({\it e.g.} ASSA).  The case of NICT,
as the sole fully deterministic forecast, appears different due to the
assignment of probabilities (see \citet{Kubo_Den_Ishii_2017} for more on
evaluation methods for fully deterministic forecasting).  This lack of
high probability forecasting is more pronounced for larger event-magnitude 
thresholds ({\it e.g.} more prevalent here for \MM\ forecasts as 
compared to \CC\ forecasts), a trend noted in \citet{allclear}.
Most of the methods here are probabilistic
with the exception of the NICT facility which produces deterministic forecasts.
Larger flares are less frequent, and probability-based
forecasts will train to reflect that fact, which reduces the presence of 
high-probability forecast values.

The ROC curves for all methods are presented in Figure\,\ref{fig:rocs},
using the same color and symbol scheme.  The $x=y$ line indicates no
ability to discriminate between the two forecast outcomes (forecast for or
against an event in the present case).  The points on the ROC curve
are computed for each distinct probability presented by a method.  
Hence, methods which provide forecasts in
discrete probability bins present with fewer points than those which
provide continuous-probability forecasts ({\it cf.} NICT {\it vs.} DAFFS).
We see a slight increase in the ability of the models that provided
forecasts for both event definitions to discriminate for the \MM\ results 
as compared to \CC.  This is a generally observed trend \citep{nci_daffs,
Murray_etal_2017}.  

Comparing the Reliability {\it vs.} the ROC plots for a particular method 
highlights the different information presented by each.  As an example, 
the MAG4 results using line-of-sight magnetograms (MAG4W, MAG4WF) {\it vs.} 
those using vector magnetograms (MAG4VW, MAG4VWF)
appear to show very similar ROC plots while displaying systematically
different behavior in the Reliability plots
(even with different training particulars with regards to longitudinal
limitations).  Also of interest are 
the comparative performances of methods which are ostensibly 
based on the same basic approaches ({\it e.g.} Poisson statistics
applied to historical region flaring rates {\it e.g.}, MCSTAT {\it vs.} ASSA --
or those with human forecasters involved -- {\it e.g.}, NOAA {\it vs.}
MOSWOC).

Figure\,\ref{fig:stats} shows the variety of skill scores and quantitative
metrics described in Section\,\ref{sec:eval}, with approximate $1\sigma$
error bars also indicated.  There is no straightforward way to estimate
uncertainties on the metrics, given the operational approach ({\it e.g.},
data for a bootstrap evaluation are not generally available).  However,
we estimate the uncertainties in two ways.  First, there are other
studies which have employed bootstrap or similar methods to calculate the
uncertainties in skill scores \citep[e.g.][]{BobraCouvidat2015,nci_daffs},
although the underlying event populations are somewhat different.
Adjusting for the smaller sample sizes here, one can estimate a general
level of uncertainty in the skill metrics of $\approx 0.06$ for \CC,
and $\approx 0.10$ for \MM.  To supplement this estimate, the DAFFS
facility (specifically the magnetic field parameter component) was re-run
for the testing interval (2016.01.01 -- 2017.12.31) using 100-draw
(with replacement) bootstrap analysis.  Across numerous metrics and
variables available in DAFFS, we find the uncertainties range over $0.04
- 0.09$ for \CC and over $0.05 - 0.17$ for \MM, with the ranges due to
whether 1- or 2-variables were tested and the particular metrics used.
These estimates are only guidance and do not necessarily reflect the
full uncertainty situation.  These uncertainties are also likely to be
underestimates, because they only account for the random error and no
separate bias is calculated for the error estimate itself.  For example
when using the full-disk bootstrap, individual days are drawn rather
than full disk-passages of individual active regions.  Additionally,
given the change in event rate between training and testing intervals,
there is likely to be a significant bias present for most methods.

The answer to the question of which methods perform ``best'' depends on
event definition and the metric under consideration.  The rank order
of performance changes between metrics and between event definitions.
This is demonstrated poignantly by MCSTAT/MCEVOL which score near
bottom rank for ApSS, but near top rank for TSS/PSS for the \MM\ tests.

Some metrics can differentiate performance better than others in these
applications.  The Proportion Correct metric for \MM\ is uninformative in
trying to differentiate between methods due to the large percentage of
correct negatives, however it provides some information for the \CC\
analysis.  Because the climatology rate does not vary across the $0.5$
threshold for \MM, the two Appleman scores (ApSS and ApSS\_clim) are identical in this case.
In the case of the \CC\ event definition, the climatology rate does
vary across the $0.5$ threshold, and the results for the two scores are
slightly different.

That being said, the majority of methods perform similarly to
each other -- that is, their scores are consistent with each other across
metrics.   This is particularly the case for the \MM\ tests given
the estimated uncertainties, although there are arguably performance
differences beyond the uncertainties for the \CC\ test.  

Comparing the Reliability plots (based on probabilities) to the Frequency 
Bias (which is a dichotomous-based metric employing a single $P_{\rm th}=0.5$)
it appears that the vast majority of methods tend toward underforecasting for larger-flare
\MM\ tests by varying degrees (FB\,$<1.0$), with a less pronounced deviation from 
FB\,$=1.0$ for most methods that underwent the \CC\ tests.  As mentioned
above, the FB score `checks' the TSS, in that for low event rates such 
as typical for solar flares, an over-forecasting system
can attain a high TSS while an under-forecasting system is less likely 
to -- so comparing TSS scores should only be performed in the context
of an accompanying FB score.  As such, for example, confidence in the TSS scores
for MCSTAT for the \MM\ test should be tempered somewhat, while
the NICT TSS result is more robust.

Different implementations of otherwise the same method
can be differentiated and the hoped-for ``improvements'' confirmed (or not).  
The implementations using vector magnetic field data 
do perform better (albeit only slightly by most metrics) than implementations
using $B_{\rm los}$ data within the same general method ({\it e.g.} MAG4W* {\it vs.}
MAG4V*, DAFFS {\it vs.} DAFFS-G).  By most metrics, MCEVOL's 
addition of an evolutionary component to MCSTAT does improve performance,
although notably not in the Gini (as visible by the shape of the ROC curve).
However, the inclusion of prior flaring history makes almost no difference
in performance across the MAG* method ({\it e.g.} MAG4W {\it vs.} MAG4WF, 
MAG4VW {\it vs.} MAG4VWF).

None of the operational methods are exceptionally good ({\it i.e.}, close to 1.0 on
any metric, except Gini and Proportion Correct), although the majority
consistently score above ``no skill'' for the metrics considered here.
Three methods demonstrate arguably poor performance specifically for
the metrics that refer to climatology; these three also show FB\,$>1.0$
(over-forecasting).  The case of NJIT is fairly well
understood and discussed below, while the others will be discussed further
in Paper III \citep{ffc3_2}.  

\section{Discussion}
\label{sec:disc}

In this study we demonstrate two things: first, a
methodology to provide meaningful
head-to-head comparisons, and second, the present state of operational
flare forecasting.  With this first direct comparison 
of forecast methods, benchmarks of performance by a
variety of measures are now provided against 
which future developments can be tested -- an important element of 
measuring progress in space weather prediction capability.

Regarding the methodology, all forecasting facilities are placed on
a level evaluation platform with respect to the full event definition
(including thresholds, validity periods and latencies).  Those whose
forecasting time differed significantly were afforded custom event
lists for evaluation, and those producing both upper- and lower-
threshold-limited forecasts were converted to exceedance forecasts to
match other methods.  Full-disk forecasts ensured that differences in
defining ``solar active regions'' would not impede the comparisons.
The time period chosen was not ideal -- too short with arguably a very
small event list -- but in the face of new data sources and a very
quiet solar cycle, it was an acceptable and necessary compromise.
Most important was how the time period was chosen -- a period that
was common to all methods which also afforded those methods relying on
SDO/HMI data an adequate training interval.

The second component of the methodology is the choice of evaluation
metrics, and this is arguably a challenge in the context of a direct
comparison because it is crucial to ensure that the metrics are
all fair (or equally unfair) to all methods.  For the presentation
here, we select a representative array of dichotomous-based and
probability-based metrics, with accompanying graphical evaluation tools,
to try and provide as complete a picture as possible.  As discussed
in \citet{allclear} and elsewhere, applying dichotomous-based metrics
to probabilistic-based forecasts require thresholds to be set which
may or may not be ideal for a particular method, resulting in unfair
penalties.  In operational practice, it is challenging to choose
the threshold that would ensure optimum performance (by measure of
various dichotomous-based metrics) at the time of forecast issuance.
As discussed in \citet{Bloomfield_etal_2012,allclear}, an optimum
threshold for TSS/PSS is usually close to the climatological event rate -- which
is itself found only after long-term averages are taken in the 
testing period.  Such information
is not available at the time of forecast issuance, and may not be optimal
for a different metric.  For the evaluations
here we encouraged methods to submit deterministic forecasts or submit
probabilistic forecasts and specify thresholds that may have been used
to produce customized deterministic forecasts for particular customers
or needs (such as an acceptable error rate of one type or the other).
None chose to provide other thresholds and thus $P_{\rm th}=0.5$ was
applied to all.  As such, we examine how well the methods perform in a
deterministic sense if action is only taken when an event is forecast
with a probability 50\% or higher.

We make note of metrics which are appropriate specifically for evaluating
operational systems, since they specifically query what value the
system brings above an available unskilled forecast.  The Appleman
and Brier skill scores by definition employ reference forecasts
based on the climatoloy of the testing period but, as discussed,
this information is not actionable for improved future performance.
We promote evaluations against an unskilled forecast.  Here we provide
analogous Mean-Square-Error Skill Score and an Appleman Skill Score
that employ a 120-day prior climatology as the reference unskilled
forecast \citep[as described in][]{SharpeMurray2017}, although others
may obviously be used.  For the testing period herein, the results
did not differ substantially from the original version of the metrics.
However, the question asked differs in a distinct way and these metrics
are highlighted as part of this work's focus on methodology.

There was not universal agreement in this group regarding evaluation
philosophy, specifically with regards to utilizing dichotomous metrics for
probabilistic forecasts.  The discussion centers on performance variation
as a function of assigned $P_{\rm th}$ in the context of an operational
system.  While a system may be trained to optimize a particular
metric and $P_{\rm th}$, there is no guarantee the performance
will be the same with that $P_{\rm th}$ during the testing interval;
evaluating a method using a new optimal $P_{\rm th}$ from the testing interval
mis-represents the performance when the information needed to assign an optimal $P_{\rm th}$
is unknown at the time of the forecast.  One approach for evaluating
probabilistic forecasts is to only employ graphical methods such as the
Reliability Plots and ROC curves and apply metrics such as the Brier
Skill Score and ROCSS (Gini score) for which no $P_{\rm th}$ is required;
this approach is fair (except to the inherently deterministic method(s))
but dismisses some metrics that the community find informative and popular.
A second approach is to present all dichotomous metrics in a manner
similar to ROC curves, displaying their outcomes as $P_{\rm th}$ is
varied and reporting the maximum attained score (with its associated
$P_{\rm th}$); but this approach can imply performance better than is
attainable in an operational setting and is unlikely to provide guidance
for improvement.  Hence, the group recognizes that the primary reason for
setting a particular $P_{\rm th}$ to apply to probabilistic forecasts is
to define a threshold upon which action should be considered according
to a particular customer's cost/benefit analysis and resilience against
forecasting errors.  The full forecast data and evaluation tools used
in the present analysis accompany this publication \footnote{Leka and Park 2019, 
Harvard Dataverse, doi:10.7910/DVN/HYP74O} so that additional metrics
using, for example, a different $P_{\rm th}$, may be calculated by the
interested reader.

Regarding the results, generally speaking no method is working
extraordinarily well; although we demonstrate that a fair number of
methods consistently perform better than various ``no skill'' measures,
meaning that they do show definitive skill across more than one metric.
No method scores above 0.5 ({\it i.e.} halfway between ``no skill'' and ``perfect'')
across all evaluation metrics, and for a number of metrics {\it no}
method provides results above 0.5.  The specific ordering of performance
varies according to metric and event definition: {\it there is no
single ``best'' method}, especially given the estimated uncertainties
in the metrics.  Amongst methods which provide different versions, the
versions generally behave similarly in some of the gross characteristics
({\it e.g.}, shapes and sampling for the ROC curves) with subtle offsets
reflecting the refinements made between each.

Three particular impacts on forecast method success are worth
noting.  First, the underlying event rate obviously varies within the
solar cycle (Fig.~\ref{fig:clims}), and possibly across solar cycles
\citep{McCloskey_etal_2018}.  This will impact the forecasting
methods, although the degree of impact will vary depending on training
methodology.  One example would be that if a method is trained
to have high reliability during a time of high solar activity, it may
then systematically over-forecast during times of declining or lower
solar activity.   Alternatively a method may not in fact be particularly
reliable during training, but when faced with a particular epoch of the
solar cycle ({\it e.g.} such as the declining phase with more isolated sunspot
groups), it may perform better.

Second, there are always flares which occur that are not assigned to
any particular active region, or occur behind the visible limb and may
be assigned to a region {\it post facto}.  During the testing period,
there were 41 unassigned \Co -flares and 3 unassigned \Mo-flares,
in some cases such unassigned flares were the sole cause of an ``event
day'' (this is discussed further in Paper IV).  Unassigned regions have
consequences for training operational systems as well as for evaluating and
testing them.  The vast majority of methods train on individual regions, and 
in doing so, they will then underforecast systematically for full-disk
forecasts.  All region-based forecasting methods will miss days where
events are produced by no assigned or detected region.

Third, we can highlight here a distinct case of the impact arising from
the lack of a full transition to operational functionality.  The NJIT method
arguably employs one of the more sophisticated analyses of magnetic
field data and shows distinct skill in the TSS and Gini metrics.
However, it arguably performs the worst according to other metrics.
Of all the methods, the NJIT system most reflects the ``research'' stage
of flare forecasting.  It was implemented without calibration across
a change in instrumentation between training and testing intervals,
which in this case (given the analysis method) could easily cause the
systematic over-forecasting as  evidenced by the metrics.  This is an issue
faced by many methods in light of aging or changing data sources and the
assumed advantage of longer training sets (see Paper III for additional
discussion on that point).  Additionally, no provisions were made for
issuing forecasts in the event of missing or delayed data, and this
severely impacted the metrics in a negative manner.  Research methods
often report encouraging results, but these must be interpreted in the
appropriate context.  In parallel, the challenge and effort required
to bring research into a fully operational mode to the point that it
is ready to undergo evaluation in an operational context must not be
underestimated.

From this presentation it is not possible to further determine 
why performances differ.  Established methods on which national
warning centers rely ({\it e.g.}, NICT {\it vs.} NOAA) display very
different characteristics in the Reliability and ROC plots, but track
fairly well amongst the evaluation metrics.  Newer methods show both
improvements and degradation against established ones ({\it e.g.}, 
MCEVOL and DAFFS {\it vs.} MOSWOC and SIDC).  However, these differences
are fairly subtle (that is, within uncertainties) when examined 
across all evaluation metrics.  

We delve further into the ``why'' question of performance differences 
in Paper III \citep{ffc3_2} by examining the impact of six distinct
categories of implementation differences, finding performance advantages to including
prior flare information and a human forecaster, and performance disadvantages
to restricting forecast-relevant data to disk-center observations.  We use a 
novel analysis method to evaluate 
temporal patterns of forecasting errors of both types ({\it i.e.}, 
misses and false alarms) for Paper IV \citep{ffc3_3}, finding weak support
for a hypothesis that including temporal information such as active region evolution 
improves a method's ability to successfully forecast, {\it e.g.}, a region's first
flare.

The obvious conclusions from this work are actually broad challenges: 
new forecasting methods, whether empirical or physics based, need to 
be evaluated {\it against these established benchmarks} with the goals 
of improved characteristics in Reliability and ROC plots, and 
metrics (specifically TSS, ApSS, ETS and BrierSS) all consistently meausuring
above 0.5 across the full range of event definitions.

\acknowledgments

We wish to acknowledge funding from the Institute for Space-Earth
Environmental Research, Nagoya University for supporting the workshop and
its participants.  We would also like to acknowledge the ``big picture''
perspective brought by Dr. M. Leila Mays during her participation in the
workshop.  S.-H.P. gratefully acknowledges Dr. Ju Jing for maintaining the
NJIT flare forecasting system and providing the archive forecasts.  KDL
and GB acknowledge that the DAFFS and DAFFS-G tools were developed under
NOAA SBIR contracts WC-133R-13-CN-0079 (Phase-I) and WC-133R-14-CN-0103
(Phase-II) with additional support from Lockheed-Martin Space Systems
contract \#4103056734 for Solar-B FPP Phase E support.  A.E.McC. was
supported by an Irish Research Council Government of Ireland Postgraduate
Scholarship. D.S.B. and M.K.G. were supported by the European Union Horizon 2020
research and innovation programme under grant agreement No. 640216
(FLARECAST project; {\tt http://flarecast.eu}).  MKG also acknowledges 
research performed under the A-EFFort project and subsequent service implementation,
supported under ESA Contract number 4000111994/14/D/ MPR.
S. A. M. is supported by the Irish Research Council Postdoctoral Fellowship Programme and
the US Air Force Office of Scientific Research award FA9550-17-1-039.
The operational Space Weather services of ROB/SIDC are partially funded
through the STCE, a  collaborative framework funded by the Belgian
Science Policy Office.  The authors thank the referees for their constructive comments.

\facilities{SDO(HMI), GONG, GOES(XRS)}
\appendix

\section{Operational Forecasting Methods: Additional Details}
\label{sec:appendix_methods}

Here we list the methods involved in the comparisons.  Pertinent details
are provided beyond the descriptions provided in the references listed
in Table\,\ref{tbl:methods}; all times here are quoted in UT.  For additional details we also suggest
referring to \citet[][Paper III]{ffc3_2}, where performance is compared
according to specific distinctions.

\subsection{A-EFFORT (Academy of Athens)}

A-EFFORT is a Space Situational Awareness (SSA) service of the
European Space Agency (ESA), available at {\tt
http://a-effort.academyofathens.gr} (with registration). Forecasts are issued at about 00:00
UT and refresh every three hours. Four exceedance thresholds are used:
M1.0+, M5.0+, X1.0+ and X5.0+, with a fixed forecast window of 24\,hr
and 0\,hr latency.

There is a single parameter computed from magnetic field
data, namely the effective connected magnetic field strength
\citep[``Beff''][]{GeorgoulisRust2007} whose values are translated
into probabilities using elements of Bayesian analysis and Laplace's
rule of succession. Beff is calculated directly up to central meridian
distances of $\pm50^\circ$; from this limit to $\pm70^\circ$ a magnetic
flux-based proxy of Beff is calculated to avoid the impact of severe
projection effects.

Each of the four forecasts is computed for each of the active regions
present within a solar meridional zone of $\pm70^\circ$, identified
using a custom active region identification algorithm \citep[see
][]{LaBonte_etal_2007}; full-disk probabilities are computed as per
Eqn.\,\ref{eqn:fd}.

\subsection{AMOS (Korean Meteorological Administration and Kyung Hee University)}

The Automatic McIntosh-based Occurrence probability of Solar activity 
(AMOS) model provides daily occurrence probabilities separately for C, 
M, and X-class flares for each NOAA active region and full disk using 
McIntosh sunspot group classes and the daily change in area for the sunspot groups.
The details are well described in \citet{amos}.

\subsection{ASAP (U. Bradford, UK)}

Described in \citet{ColakQahwaji2008,ColakQahwaji2009}, ASAP also participated 
in the ``All Clear'' workshop in 2009 \citep{allclear}.

\subsection{ASSA (Korean Space Weather Center)}

The Automatic Solar Synoptic Analyzer (ASSA) system at the Korean
Space Weather Center identifies and predicts for a variety of solar
activity, including sunspot groups and associated flaring. Flare
forecast results are issued hourly at :00, with a McIntosh-class-based
forecast extending for 24h (used here, initiated in late 2013) and a new
``parameter-based'' forecast using six major parameters extending for
12h. The McIntosh-class-based forecast uses an independent ASSA algorithm
(not NOAA determinations) to identify sunspot groups and determines
their McIntosh class by estimating their morphological characteristics,
and produces an independent flaring probability according to the ASSA
sunspot-flare archive (not based on otherwise published rates). The
ASSA sunspot-flare archive was produced based on statistical matching
between ASSA's sunspot group catalog and NOAA's GOES Soft X-ray events catalog
during 1996--2013.  A parameter-based method was initiated in late 2016,
and provides flare forecasts based on multi-component linear regression
using parameters such as the number of sunspots in a sunspot group, the
total area of sunspots in a group, and the group's longitudinal extent.
Unfortunately, forecasts from this second method were not submitted.
ASSA forecasts rely on SDO/HMI continuum and line-of-sight magnetogram
images with no correction for limb-ward effects.  Additional details
may be found in the user manual \citep{ASSA_DOC}.

\subsection{BOM (Flarecast, Bureau of Meteorology, Australia)}

The details of the probabilistic model are well described
in \citet{Steward_etal_2011,Steward_etal_2017}.  Flarecast II
(not yet published but results are submitted here) uses the SDO 
HMI magnetogram imagery analysis
capability developed for the original Flarecast model
\citep{Steward_etal_2017} plus prior flaring history, and adds a machine
learning technique (logistic regression) to generate a probabilistic
forecast.  Variables that describe HMI $B_{\rm los}$ magnetograms are
selected to minimize Aikake's Information Criteria (AIC), and logistic
regression is used to estimate the coefficients of the model, and
then used to generate M+, X+, region and full-disk, probabilistic and
categorical deterministic forecasts output for flaring activity over the next
24 hours.  In operational mode the predictions are updated at 00, 06,
12, 18 UT.

\subsection{DAFFS, DAFFS-G (Discriminant Analysis Flare Forecasting System, NorthWest Research Associates (NWRA))}

{\tt DAFFS} is well described in \citet{nci_daffs}, but it should be noted
that it is a fairly young, recently-released system.  Of note, being
the only method to primarily rely on quantitative analysis of vector
magnetic field data from a non-operational data source ({\it SDO}/HMI), this
method suffered from data problems arising from the data-acquisition
mode change that incurred a temporary data 
mis-alignment\footnote{see \tt{http://hmi.stanford.edu/hminuggets/?p=1596} and the 
{\it SolarNews} note of 01 September 2017 at \tt{https://solarnews.nso.edu/2017.html\#20170901}.}
(MAG4V* methods use {\it SDO}/HMI data in a more limited fashion, see below).  The impacted
data spanned April 2016 -- September 2017, and was most damaging for
data away from disk center.  (The ``definitive'' data have subsequently
been re-processed; the ``near real time'' data will not be).  We noted
that it most dramatically impacted some parameters in top-performing
combinations, but not others.  For the results here, we modified
DAFFS to run using parameter combinations that performed essentially
identically (within the metric error bars) in the training phase but were
not as susceptible to the HMI vector data problem: specifically for the
\CC\ event definition the parameter combination was changed to
$[E_e,\log(\mathcal{R_{\rm nwra}})]$ and the \MM\ event definition
parameter pair was changed to $[{\rm FL}_{\rm 24}, \log(\mathcal{R_{\rm
nwra}})]$ from what is described in \citet{nci_daffs}.  

The {\tt DAFFS-G} ( tool runs simultaneously and is based primarily on GONG
$B_{\rm los}$ data and persistence (NOAA near real time (NRT) event reports).   {\tt DAFFS-G}
is a very ``young'' release, and has not yet been fully optimized for performance.
For the forecasts submitted here, the parameter combinations were 
$[\,\overline{\nabla(B_z^{\rm pot})}, \Phi^{\rm pot}_{\rm tot}]$ for \CC, and
the parameters for \MM\ were $[\sigma(\nabla(B_h^{\rm pot})), \Phi^{\rm pot}_{\rm tot}]$,
where the ``pot'' moniker refers to the potential field calculated from the $B_{\rm los}$
data \citep{bbpot}.

\subsection{MAG4* (NASA/Marshall Space Flight Center)}

MAG4 is described in \citep{Falconer_etal_2011,Falconer_etal_2014}.   
This study included four 
versions:
\begin{itemize}
\item MAG4W: Free-energy Proxy Only using Line-of-Sight Magnetogram
\item MAG4WF: Free-energy Proxy and Previous Flare History using Line-of-Sight Magnetograms
\item MAG4VW:  Free-energy Proxy Only using Deprojected HMI Vector Magnetogram
\item MAG4VWF   Free-energy Proxy and Previous Flare History using Deprojected HMI Vector
    Magnetograms
\end{itemize}

MAG4W[F] uses the HMI NRT $B_{\rm los}$ data with no further correction. 
The MAG4VW and MAG4VWF, like DAFFS, use SDO/HMI vector magnetic field data, however only
to $30^\circ$ from disk center, which were minimally impacted by the data misalignment.
In MAG4*F, previous flare information is used, although a region is assumed to be 
non-flaring if that information is not available.  

\subsection{MCSTAT, MCEVOL (MaxMillenium Flare Prediction System)}

The MCSTAT approach is well described in  \citet{gallagheretal02,Bloomfield_etal_2012} 
while the MCEVOL approach is well described in \citet{McCloskey_etal_2018}.

\subsection{MetOffice (UK) MOSWOC}

The details are well described by \citet{Murray_etal_2017}.  Of note, the
forecast closest to 00:00 was used, but is not necessarily the official
forecast for that day from MOSWOC, as updates are applied through the
(local) night.

\subsection{NICT (National Institute of Information and Communications Technology, Japan)}

The details of this long-running system are well described in
\citet{Kubo_Den_Ishii_2017}.  Unique to the methods, the NICT-human
approach provides four categorical deterministic forecasts of maximum
flare size: ``Quiet'' (max: {\tt A/B}-class), ``Eruptive'' (max: {\tt
C-}class), ``Active'' (max: {\tt M-}class) or ``Major Flare'' (max: {\tt
X-}class).  These were converted to probabilities of $[0.0,~1.0]$ for
the probabilistic-based analysis and converted to exceedance forecasts.

\subsection{NJIT (New Jersey Institute of Technology)}

The basic methodology is described in \citet{Park_Chae_Wang_2010}.
The NJIT method is operational in the sense it produces forecasts
automatically, but has not been developed further since 2010. It provides
probabilistic forecasts of at least one {\tt C-}, {\tt M-} and {\tt
X-}class flare occurrence only for a given NOAA-numbered active region
within $\pm60^\circ$ of disk center; these were converted to 
exceedance forecasts.  The method was trained on 300
primarily flare-productive active regions using SOHO/MDI line-of-sight
active-region magnetic field data in solar cycle 23. However, the
forecasts now use HMI line-of-sight data without any cross-calibration
between the two data sources.

\subsection{NOAA (Space Weather Prediction Center, US National Oceanic and Atmospheric 
Administration NOAA)}

The forecasts by NOAA/SWPC have long been considered a standard 
\citep{Crown2012} and have set the benchmarks against which methods
are measured using the NOAA/SWPC event definitions (see commentary
on this in \citet{EEG_chapter}).  SWPC forecasters begin with a climatological
basis according to an active region's classification (SWPC's assignment
of active region class is also considered ``The Standard''), according
to the historical flaring rates of different sunspot region classes
\citep{McIntosh1990}.  From this, a forecaster may modify
a region's probability according to region evolution, flaring trends,
and forecaster experience and expertise.   These region probability
forecasts are combined for a full-disk forecast which itself may be
modified based on flaring history of recently-rotated-off regions, or
indications of a highly active region about to return.  Forecasters may
also incorporate other model data when available.  Initial forecasts are
issued at 22:00 (the ``Geophysical Activity Report and Forecast'' or RSGA)
valid beginning at 00:00 the next day.  These are incorporated into the ``3-day Forecast''
issued at 00:30, with a minimal but not zero probability of a forecast update
in the intervening 2.5\,hr.  Forecasts can, but are not likely to, be updated 
again before the next 3-day forecast is issued at 12:30.  The data used
in this comparison arise from the 3-day forecasts but include the \CC\ 
forecasts that are not generally published.

\subsection{SIDC (Solar Influence Data Analysis Centre of the Royal Observatory of Belgium)}

The forecaster on duty at the SIDC produces each day (nominal issue
time 12:30UT) a probabilistic forecast for the occurrence of X-ray
flares over the next 24h. Probabilities are provided for flare classes
{\tt C-}, {\tt M-} and {\tt X-} separately. A full disk as well as
an active region specific forecast is provided.  The forecasters use
various data sources, the main one being the flaring probability from
active regions with the same McIntosh classification. Such probability
is then modulated using for example: the specific flare histories for
the regions to be forecasted, {\it SDO}/HMI magnetogram movies, {\it
SDO}/AIA movies, and {\it STEREO}/EUVI movies {\it e.g.} to assess the
flaring activity of active regions rotating onto or off the solar disk.
Details on flare forecasting at ROB/SIDC and its validation procedures
are provided in~\cite{Berghmans_etal_2005,Devos_etal_2014}.

\section{Steps to Produce Full-Disk Exceedance Forecasts.}

\subsection{Full-Disk Forecasts from Region-Based Forecasts}
\label{sec:appendix_FD}

The forecasts considered here are ``full-disk'' forecasts, meaning essentially
treating the Sun as a star.  In practice, only one method did not produce full-disk forecasts, meaning
that they only provided forecasts for active regions individually. In that case, the
region probabilities were combined according to,
\begin{equation}
\label{eqn:fd}
P_{\rm FD} = 1.0 - \Pi_{\rm AR}(1.0 - P_{\rm AR})
\end{equation}
where $P_{\rm AR}$ is the probability of an event for each active region, and
the product is performed over all active regions for which such a probability is 
provided.  This equation is effectively how all region-forecasting methods produce 
their baseline full-disk forecasts.

\subsection{Class-Specific {\it vs.} Exceedance Forecasts}
\label{sec:appendix_combos}

The results from methods producing class-specific forecasts ({\it e.g.}
M1.0 - M9.9) were converted to exceedance forecasts ({\it e.g.} M1.0+
with no upper limit) using conditional probabilities over that method's
training interval, by the following methodology.  Suppose one has the
probabilities of occurrence of at least one {\tt C-}, {\tt M-} and {\tt
X-} class flares respectively, for a given forecast time window $\tau$
denoted by $P$(C) for \Co, $P$(M) for \Mo\ and
$P$($\geq$X1)=$P$(X) for \Xp. Then, the lower-bound
only probabilities of $P$($\geq$C) and $P$($\geq$M) can be determined
by combining the probabilities of $P$(C), $P$(M) and $P$($\geq$X1)
with their associated conditional probabilities.

The probability of occurrence of at least one flare at the level greater
than or equal to M1.0 during $\tau$, i.e., $P$($\geq$M1), can be derived
as follows,

\begin{eqnarray}
\label{eqn:combo1} 
P\mathrm{({\geq}M1)} &=& P\mathrm{(M)} + P\mathrm{(X)} - P\mathrm{(M\,\,and\,\,X)} \nonumber \\
&=& P\mathrm{(M)} + P\mathrm{(X)} - P\mathrm{(M)} \times P\mathrm{(X|M)},
\end{eqnarray}
where $P$(M and X) is the probability that both M- and X-class flares will occur at least once during $\tau$, and $P$(X$|$M) is the conditional probability of at least one X-class flare occurring given at least one M-class flare occurred during $\tau$.

Similarly, $P$($\geq$C1) can be determined as follows:
\begin{eqnarray}
\label{eqn:combo2} 
P\mathrm{({\geq}C1)} &=& P\mathrm{(C)} + P\mathrm{(M)} + P\mathrm{(X)} - P\mathrm{(C\,\,and\,\,M)} - P\mathrm{(C\,\,and\,\,X)} \nonumber \\
&& - P\mathrm{(M\,\,and\,\,X)} + P\mathrm{(C\,\,and\,\,M\,\,and\,\,X)} \nonumber \\
&=& P\mathrm{(C)} + P\mathrm{(M)} + P\mathrm{(X)} - P\mathrm{(C)} \times P\mathrm{(M|C)} - P\mathrm{(C)} \times P\mathrm{(X|C)} \nonumber \\
&& - P\mathrm{(M)} \times P\mathrm{(X|M)} +  P\mathrm{(C)} \times P\mathrm{(M|C)} \times P\mathrm{(X|C\,\,and\,\,M)},
\end{eqnarray}
where $P$(X$|$C and M) is the conditional probability of at least one
X-class flare occurring given both {\tt C}- and {\tt M}-class flares occurred at
least once during $\tau$.

The conditional probabilities are calculated using the NOAA/SWPC
historical flare event list data and $\tau$ as the prescribed
validity interval ({\it e.g.}, 24\,hr) starting from 00:00 UT of a
given date. In this case, for example, $P$(X$|$M) can be determined
as follows: 
\begin{enumerate} 
\item During the training interval for a given forecast method, we find the 
dates $D$(M) on which at least one {\tt M}-class flare occurred.  
\item From the dates $D$(M), we determine the subset $D$(X$|$M) of dates on which 
at least one flare at the level greater than or equal to {\tt X1.0} occurred.
\item The conditional probability
$P$(X$|$M) is then the total number of elements in $D$(X$|$M) divided by the total
number of $D$(M).  
\end{enumerate}

\begin{figure}[h]
\centering
\includegraphics[width=10cm]{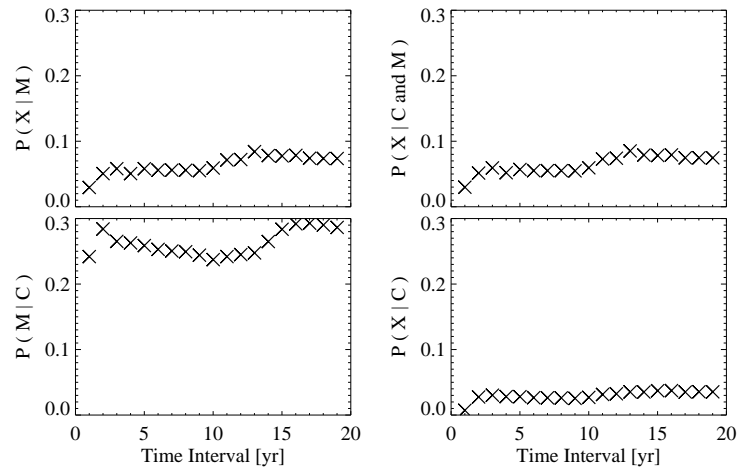}
\caption{{\tt C-}, {\tt M-}, {\tt X-}class flare conditional probabilities 
as function of different time intervals as used for the
calculations of exceedance.  Time interval extends {\it back}
in time from 2015.12.31.}
\label{fig:condprobs}
\end{figure}

The other conditional probabilities can be calculated in the
same way as $P$(X$|$M) explained above.  Figure\,\ref{fig:condprobs} shows the
conditional probabilities for different time intervals used for their
calculations. Note that the end date of all of the time intervals is
fixed at 23:59 UT on 2017-Dec-31.  The conditional
probabilities do not significantly change as a function of the time interval. 
Because our goal is to calculate $P$($\geq$C1) and $P$($\geq$M1)
from the probabilities of $P$(C), $P$(M) and $P$($\geq$X1) that a
given forecast method provides, the proper time interval to use for
calculating the conditional probabilities is the training interval
for that specific forecast method.

Forecasts for flare-class specific probabilities are converted to 
exceedence forecasts for the following methods: AMOS, ASAP, ASSA,
MOSWOC, NICT, and NJIT.

\end{document}